\documentclass[journal]{IEEEtran}

\usepackage{amsmath,graphicx}
\usepackage{amsfonts} 
\usepackage{mathtools}
\usepackage{pdfpages}
\usepackage{booktabs}
\usepackage{tabularray}
\usepackage{comment}
\usepackage{xcolor}
\usepackage{hyperref}
\ifCLASSINFOpdf
\else
\fi
\begin{document}
\title{Bridging The Multi-Modality Gaps of Audio, Visual and Linguistic for Speech Enhancement}

\author{Meng-Ping Lin, 
        Jen-Cheng Hou, 
        Chia-Wei Chen,
        Shao-Yi Chien, 
        Jun-Cheng Chen,
         \\Xugang Lu, 
        Yu Tsao,~\IEEEmembership{Senior Member,~IEEE,}

\thanks{Meng-Ping Lin, Chia-Wei Chen, and Shao-Yi Chien are with National Taiwan University, Taiwan, corresponding email: (mplin@media.ee.ntu.edu.tw; cwchen@media.ee.ntu.edu.tw; sychien@media.ee.ntu.edu.tw). Shao-Yi Chien is the corresponding author.}

\thanks{Jen-Cheng Hou, Jun-Cheng Chen, and Yu Tsao are with Academia Sinica, Taipei, Taiwan, corresponding email: (jchou@citi.sinica.edu.tw; pullpull@citi.sinica.edu.tw; yu.tsao@citi.sinica.edu.tw).}

\thanks{Xugang Lu is with National Institute of Information and Communications Technology, Japan, corresponding email: (xugang.lu@nict.go.jp).}}

\maketitle

\begin{abstract}
Speech enhancement (SE) aims to improve the quality and intelligibility of speech in noisy environments. Recent studies have shown that incorporating visual cues in audio signal processing can enhance SE performance. Given that human speech communication naturally involves audio, visual, and linguistic modalities, it is reasonable to expect additional improvements by integrating linguistic information. However, effectively bridging these modality gaps—particularly during knowledge transfer remains a significant challenge. In this paper, we propose a novel multi-modal learning framework, termed DLAV-SE, which leverages a diffusion-based model integrating audio, visual, and linguistic information for audio-visual speech enhancement (AVSE). Within this framework, the linguistic modality is modeled using a pretrained language model (PLM), which transfers linguistic knowledge to the audio-visual domain through a cross-modal knowledge transfer (CMKT) mechanism during training. After training, the PLM is no longer required at inference, as its knowledge is embedded into the AVSE model through the CMKT process. We conduct a series of SE experiments to evaluate the effectiveness of our approach. Results show that the proposed DLAV-SE system significantly improves speech quality and reduces generative artifacts—such as phonetic confusion—compared to state-of-the-art (SOTA) methods. Furthermore, visualization analyses confirm that the CMKT method enhances the generation quality of the AVSE outputs. These findings highlight both the promise of diffusion-based methods for advancing AVSE and the value of incorporating linguistic information to further improve system performance.
 \\
\end{abstract}

\begin{IEEEkeywords}
%IEEE, IEEEtran, journal, \LaTeX, paper, template.
diffusion model, predictive and generative model, audio-visual speech enhancement (AVSE), pretrained language model (PLM), cross-modal knowledge transfer.
\end{IEEEkeywords}

% \fi
\IEEEpeerreviewmaketitle

\section{Introduction}
%\IEEEPARstart{T}{his} demo file is intended to serve as a ``starter file''
%for IEEE journal papers produced under \LaTeX\ using
%IEEEtran.cls version 1.8b and later.
%I wish you the best of success.

 

\IEEEPARstart{D}{eep} learning (DL) has emerged as a powerful approach in speech enhancement (SE), leading to significant progress in the field \cite{OverView}. By leveraging deep architectures, DL models can learn complex representations from noisy speech signals, enabling the effective reconstruction of clean speech. Various DL-based models have been successfully applied to SE, including deep denoising autoencoders (DDAE) \cite{DDAE1}, fully connected neural networks \cite{FCNN2, FCNN3}, convolutional neural networks (CNN) \cite{CNN1}, recurrent neural networks (RNN) and long short-term memory (LSTM) \cite{RNN6}, attention mechanism \cite{TFAttn, TFAttn2}, and hybrid architectures that combine multiple neural-network architectures \cite{CRNSE}. These DL-based approaches consistently outperform traditional SE techniques and earlier statistical or machine learning methods \cite{PESQNet, Phase, V2V, V2V2}.

A key advantage of DL models is their ability to integrate information from multiple domains. One prominent direction in this area is audio-visual speech enhancement (AVSE), which incorporates visual cues to improve SE systems. Traditional audio SE systems primarily rely on audio signals alone, which can limit performance in adverse acoustic conditions such as non-stationary noise, reverberation, or the presence of competing speakers. To address these limitations, the AVSE approach \cite{AVSE1, AVSE7, AVSE9} has emerged as a promising direction by incorporating visual cues—especially lip movements and facial expressions—that are inherently correlated with speech production. Visual information is immune to acoustic noise and can provide complementary features when the auditory signal is degraded. This cross-modal synergy has been shown to significantly improve the SE performance, especially in low signal-to-noise ratio (SNR) conditions.

Numerous AVSE and audio-visual speech separation (AVSS) systems have been proposed and have demonstrated strong performance \cite{AVSE10, Conversion, AVSE4}. Compared to audio-only SE systems, AVSE requires additional visual inputs and model parameters to process the visual information. To reduce complexity, some approaches focus specifically on the lip region rather than processing the entire face \cite{AVSE3, AVSE5}. In addition, a variety of input types \cite{AVSE2, AVSS_FSM} and model architectures \cite{AVSE6,  AVSE_Haizhou, AVSE_DeLiang, AVSE11, AVSE_Kalman} have been explored for AVSE. Furthermore, unified learning frameworks have been introduced to jointly train AVSE with auxiliary tasks, improving robustness and generalization \cite{Joint_AVSE, AVSE8}.

Recently, diffusion models \cite{Diffusion} have gained significant attention for their impressive performance in various vision and speech generation tasks. In the SE field, diffusion models have also been adopted as core architectures. For example, \cite{DSE_WGS} proposed a diffusion-based SE model that combines a supervised loss with a generative Gaussian noise prediction loss, demonstrating outstanding SE performance. Similarly, \cite{Seeing} introduced an audio-visual speech separation (AVSS) system that leverages a two-stage diffusion process—consisting of a predictive stage followed by a generative stage—achieving state-of-the-art (SOTA) results on multiple benchmarks. Several other diffusion-based SE systems have also shown strong effectiveness \cite{AV2Wav, Ensemble}.

Speech communication is inherently multimodal, involving audio, visual, and linguistic cues. While AVSE has demonstrated superior performance over audio-only approaches, integrating linguistic information offers additional advantages—especially when audio or visual inputs are unreliable or degraded. Linguistic knowledge provides context-aware guidance based on language structure and semantics, enabling AVSE models to better disambiguate overlapping or corrupted speech. This leads to improved intelligibility, more accurate reconstruction, and reduced artifacts. By aligning noisy inputs with plausible linguistic patterns, AVSE models can more effectively isolate the target speaker. Overall, the integration of linguistic information enhances the robustness and effectiveness of multimodal SE systems.

In this study, we propose a novel DLAV-SE system, a diffusion-model-based framework that jointly incorporates audio, visual, and linguistic modalities. Both the audio and visual components utilize a two-stage process comprising a predictive stage and a generative stage, designed to reconstruct clean speech from noisy audio and lip movements, respectively. To integrate linguistic knowledge, we introduce a cross-modal knowledge transfer (CMKT) mechanism, which leverages context-dependent representations from a pretrained BERT model. This linguistic module guides the training of the AVSE model through linguistically informed loss functions, improving its ability to produce clear and contextually appropriate speech. For efficient linguistic knowledge transfer learning, we designed two types of alignment and matching methods, i.e., Multi-Head Cross-Attention (MHCA) \cite{Transformer} and Optimal Transport (OT) \cite{VillanoBook}. Both methods facilitate the transfer of linguistic knowledge across audio-visual and linguistic modalities by bridging domain gaps. Since the linguistic framework is used only during training, it adds virtually no inference cost, demonstrating parameter and computational efficiency.

To evaluate the effectiveness of the proposed DLAV-SE system, we conducted experiments on three datasets: a Mandarin dataset (Taiwan Mandarin Speech with Video, referred to as TMSV), an English dataset (the 3rd AVSE Challenge dataset, referred to as AVSEC-3), and a real-world in-vehicle speech dataset recorded on the Academia Sinica campus, named AS-CarSpeech. Experimental results demonstrate that DLAV-SE consistently outperforms existing methods across both language settings. Furthermore, our findings confirm the benefits of incorporating linguistic information, showing significant performance gains in Mandarin, English, and real-world AVSE tasks.

\begin{figure}[tb]
	\centering
    \includegraphics[scale=0.4]{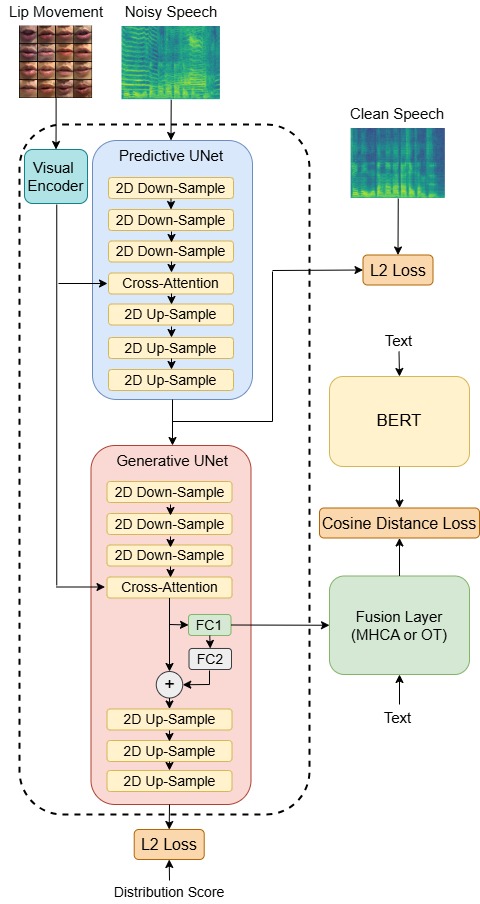}
    \caption{The hybrid predictive–generative diffusion-based DLAV-SE system is trained with integrated linguistic information. The linguistic module, including the BERT model, cosine distance loss, and fusion layer, is utilized only during the training phase. During inference, only the components enclosed by the dotted line are activated.}
	%\vspace{-4mm}
    \label{fig:training_framework}
\end{figure}

\section{Related Works}
Recent advances have established a theoretical connection between diffusion models and score matching through the analysis of stochastic differential equations (SDEs) derived from discrete-time Markov chains \cite{ScoreDiff, ScoreMatch}. This relationship allows the forward diffusion process to be reversed via a corresponding reverse-time SDE, which depends solely on the score function of the perturbed data \cite{Perturbed}. Building on this foundation, several studies—including \cite{SGMSE, ScoreAVSE}—have proposed SE systems based on score-based diffusion models, where both the forward and reverse processes are formulated using SDEs. These works extend the theoretical framework introduced in \cite{ScoreDiff}. In our study, we adopt the score-based diffusion model and the feature preprocessing techniques from \cite{SGMSE} as the foundation for our model architecture.

In parallel with advances in diffusion modeling, recent studies have demonstrated the effectiveness of PLMs in transferring context-dependent linguistic knowledge to enhance the performance of audio models in speech classification tasks such as automatic speech recognition (ASR). This transfer mechanism is commonly referred to as CMKT. For instance, \cite{LuCTCOP} leveraged a PLM with OT to improve an end-to-end connectionist temporal classification (CTC)-based ASR system, while \cite{LuCTCSink} introduced a hierarchical framework incorporating Sinkhorn attention \cite{SinkAttn} to further boost the performance of a CTC-OT-based model. Inspired by these developments, we investigate the application of CMKT to AVSE, incorporating it during the training phase of our proposed system to bridge domain gaps between modalities and enable effective linguistic knowledge integration.

To benchmark our proposed system, we review several existing AVSE models evaluated on the TMSV dataset. AVDCNN \cite{AVSE1} integrates audio and visual streams within a unified network and adopts a multi-task learning framework for AVSE. iLAVSE \cite{ImprovedLite} employs a convolutional recurrent neural network architecture and demonstrates the effectiveness of visual input, specifically focusing on the mouth region of interest (ROI) to improve efficiency over using full-face features. AVCVAE \cite{AVSE6} introduces audio-visual extensions of variational autoencoders (VAEs) for single-channel, speaker-independent SE. SSL-AVSE \cite{SSLAVSE} leverages self-supervised learning by utilizing AV-HuBERT \cite{AVHubert} to fuse visual and audio signals. DCUC-Net \cite{DCUC-Net} features a complex U-Net architecture designed to process complex-domain features using stacked conformer blocks. 

For the AVSEC-3 English dataset, we compare the proposed DLAV-SE system against several competitive baselines from the 3rd AVSE Challenge. MMDTN \cite{MMDTN} introduces a multi-modal dual-transformer architecture that captures cross-modal correlations through attention mechanisms. LSTMSE-Net \cite{LSTMSE-Net} uses a LSTM network to process audio and visual features via a specialized separator module for improved AVSE performance. These systems provide strong benchmarks for evaluating the effectiveness of our proposed approach.
  
\begin{figure*}[tb]
	\centering
    \includegraphics[scale=0.4]{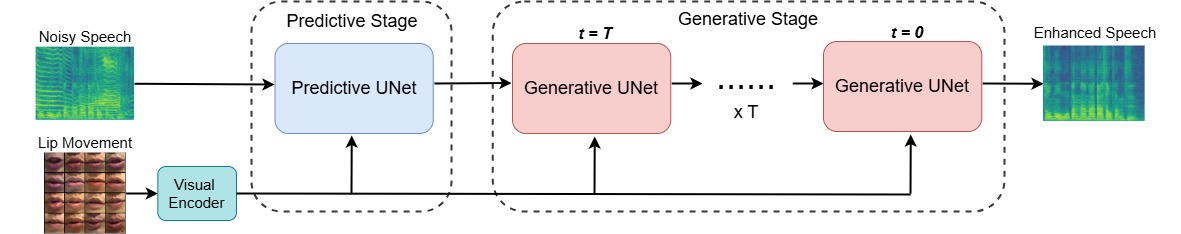}
    \caption{The inference process of the DLAV-SE System. After applying the short-time Fourier transform (STFT), the input noisy spectrogram is first passed through the predictive model to generate a preliminary denoised output. This intermediate spectrogram is then fed into the generative model, which performs reverse diffusion over T steps. Each diffusion step is implemented using a U-Net architecture \cite{UNet}.}
    \label{fig:Diffusion}
\end{figure*}

\section{Proposed method}
\label{sec:proposed}
% Two modalities
Our model consists of two main branches, as illustrated in the overall DLAV-SE training architecture in Fig.~\ref{fig:training_framework}. The left branch builds upon the diffusion-based audio-visual separation system "DAVSE" \cite{DAVSE}, which we adopt and adapt for our purposes. This work extends the use of diffusion models \cite{CondDiff2022} to the audio-visual domain by incorporating visual information directly into the diffusion process.

The right branch processes the linguistic modality, incorporating a pretrained language model (BERT) \cite{BERT} followed by a fusion layer. Notably, this linguistic branch is utilized only during training and excluded from the inference phase. The complete diffusion process is illustrated in Fig.~\ref{fig:Diffusion}.

% -----------------------Figure:Two modalities-----------------------%

% ++++++++++++++++++++++++++++ViusalModality++++++++++++++++++++++++
% Model Architecture
\subsection{Lip Movement Model}
%\hspace{3ex}
To improve the relevance and efficiency of visual information in DLAV-SE, we focus on the mouth region, rather than the full face, as the region of interest (ROI). Lip movements offer the most discriminative visual cues for speech, enabling the visual model to concentrate on the most informative area and thereby enhance the overall AVSE performance.

For visual feature extraction, we use a modified ResNet-18 encoder \cite{Lip} pretrained on the Lip Reading in the Wild (LRW) dataset. The encoder begins with a 3D convolutional layer that captures temporal dynamics from sequences of consecutive frames, followed by a 2D ResNet-18 that extracts spatial features. The resulting visual embeddings are then fed into the downstream predictive and generative models.

The visual encoder is pretrained and frozen during training, allowing us to precompute and store visual embeddings for all datasets. This design choice eliminates the need to include the visual model during training, significantly reducing the number of trainable parameters and overall memory consumption.
\begin{equation}
	\begin{array}{l}				
		\mathbf{v} = \mathbf{Encoder} \left( {\bf P} \right), \\
	\end{array}
	\label{eq:visual_encoder}
\end{equation}
where ${\bf Encoder}$ is the lip movement model, ${\bf v}$ is the visual embedding vectors, and ${\bf P}$ is the ROI as the input matrix. 

% ++++++++++++++++++++++++++++++AudioModality+++++++++++++++++++++++++++
% Model Architecture
\subsection{Predictive Denoiser Model}
To initialize the diffusion process, we introduce a predictive denoiser. As the starting point strongly affects the performance of the generative model, providing an accurate initial noise estimate can substantially improve the final output quality. The predictive denoiser receives a spectrogram of the noisy speech, which is obtained via the STFT, and produces an initial noise estimate. This estimate is then fed into the generative diffusion model to guide the denoising process. By offering a more informed initialization, the predictive module helps steer the diffusion process toward more precise and effective SE performance.

We adopt the same model architecture for the predictive denoiser as that used for the generative score model, which will be detailed in a later section.
\begin{equation}
	\begin{array}{l}
		%\begin{align}					
		\mathbf{\hat{y}} = \mathbf{Denoiser} \left( {\bf v}, {\bf y} \right), \\
		%\end{align}
	\end{array}
	\label{eq:denoiser}
\end{equation}
where ${\bf Denoiser}$ is the predictive denoiser, ${\bf y}$ is the noisy speech spectrogram and ${\bf \hat{y}}$ is the preliminary-denoised speech spectrogram. 

To train this denoiser, we apply a mean square error (MSE) loss between the denoised speech spectrogram and the clean speech spectrogram.
\begin{equation}
	\begin{array}{l}
		%\begin{align}					
		{L_{denoiser}} = {\bf MSE} \left( {\bf x}, {\bf \hat{y}} \right), \\
		%\end{align}
	\end{array}
	\label{eq:denoiser_loss}
\end{equation}
where ${L_{denoiser}}$ is the loss of the predictive denoiser, ${\bf x}$ is the clean speech spectrogram. 
\subsection{Score-Based Generative Modeling}

To model the audio modality in the proposed SE framework, we adopt a score-based diffusion model called ``SGMSE,'' proposed by Richter et al. \cite{SGMSE}. This model operates in the complex STFT domain and leverages SDEs to model the generative process of speech spectrograms. In the forward process, noise is gradually added to a clean speech spectrogram, governed by the following SDE: 

\begin{equation}
	\begin{array}{l}
		{\bf d{x}}_t = {\bf f} \left( {\bf x}_t, {\bf \hat{y}} \right){\bf d}t + {\bf g}{(t)}{\bf dw},
	\end{array}
	\label{eq:forward}
\end{equation}
where ${\bf x}_t$ is the noisy spectrogram at time $t \in [0, T]$, ${\bf \hat{y}}$ is the preliminary-denoised spectrogram obtained from the predictive model, and ${\bf w}$ is a standard Wiener process.

The drift and diffusion coefficients are defined as: 

\begin{equation}
	\begin{array}{l}
		{\bf f} \left( {\bf x}_t, {\bf \hat{y}} \right) = {\bf \eta} \left( {\bf \hat{y}} - {\bf x}_t \right),
	\end{array}
	\label{eq:drift_coefficient}
\end{equation}
\begin{equation}
	\begin{array}{l}
		{\bf g}{(t)} = {\bf \sigma}_{\text{min}} \left( \dfrac{{\bf \sigma}_{\text{max}}}{{\bf \sigma}_{\text{min}}} \right)^t \sqrt{2 \log \left( \dfrac{{\bf \sigma}_{\text{max}}}{{\bf \sigma}_{\text{min}}} \right)},
	\end{array}
	\label{eq:diffusion_coefficient}
\end{equation}
where $\eta$ is a stiffness constant that regulates the transition from ${\bf x}_0$ to ${\bf \hat{y}}$, and ${\bf \sigma}_{\text{min}}$, ${\bf \sigma}_{\text{max}}$ define the noise schedule.

The reverse process generates the clean spectrogram by iteratively denoising from ${\bf x}_T$ to ${\bf x}_0$:
\begin{equation}
	\begin{array}{l}
		{\bf d{x}}_t = \left[ -{\bf f} \left( {\bf x}_t, {\bf \hat{y}} \right) + {\bf g}{(t)}^2 \nabla_{{\bf x}_t} \log p_t({\bf x}_t|{\bf \hat{y}}) \right]{\bf d}{t} + {\bf g}{(t)}{\bf d}{\bf \overline{w}},
	\end{array}
	\label{eq:reverse}
\end{equation}

where the score function $\nabla_{{\bf x}_t} \log p_t({\bf x}_t|{\bf \hat{y}})$ is estimated by a deep neural network (DNN), denoted as ${\bf s}_\theta({\bf x}_t, {\bf \hat{y}}, t)$. This leads to the following reformulated SDE:
\begin{equation}
	\begin{array}{l}
		{\bf d{x}}_t = \left[ -{\bf f}({\bf x}_t, {\bf \hat{y}}) + {\bf g}{(t)}^2 {\bf s}_\theta({\bf x}_t, {\bf \hat{y}}, {\bf v}, t) \right] {\bf d}{t} + {\bf g}{(t)} {\bf d} {\bf \overline{w}}.
	\end{array}
	\label{eq:reverse_score}
\end{equation}

At inference, the initial state ${\bf x}_T$ is sampled as:
\begin{equation}
	\begin{array}{l}
		{\bf x}_T \sim \mathcal{N}_C \left( {\bf \hat{y}}, \sigma(T)^2 {\bf I} \right),
	\end{array}
	\label{eq:Initial_Condition}
\end{equation}
where $\mathcal{N}_C$ is a circularly symmetric complex Gaussian distribution, and ${\bf I}$ is the identity matrix. 
The enhanced spectrogram is then converted to a waveform using inverse STFT (ISTFT).

During training, we pass sampled triples $({\bf x}_t, {\bf \hat{y}}, t)$ into the score model using the following steps:
1. Sample $t$; 
2. Sample $({\bf x}_0, {\bf \hat{y}})$ from the dataset;
3. Sample $\boldsymbol{\phi} \sim \mathcal{N}_C(\mathbf{0}, \mathbf{I})$;
4. Generate ${\bf x}_t$ using:
\begin{equation}
	\begin{array}{l}
		{\bf x}_t = \mu({\bf x}_0, {\bf \hat{y}}, t) + \sigma(t)\boldsymbol{\phi},
	\end{array}
	\label{eq:Sample_Xt}
\end{equation}
where:
\begin{equation}
	\mu({\bf x}_0, {\bf \hat{y}}, t) = e^{-{\eta}t}{\bf x}_0 + (1 - e^{-{\eta}t}){\bf \hat{y}},
	\label{eq:Gaussian_Mean}
\end{equation}
\begin{equation}
	\sigma(t)^2 = \dfrac{{\sigma}_{\text{min}}^2 \left( \left( \dfrac{{\sigma}_{\text{max}}}{{\sigma}_{\text{min}}} \right)^{2t} - e^{-2\eta t} \right) \log \left( \dfrac{{\sigma}_{\text{max}}}{{\sigma}_{\text{min}}} \right)}{\eta + \log \left( \dfrac{{\sigma}_{\text{max}}}{{\sigma}_{\text{min}}} \right)}.
	\label{eq:Gaussian_Variance}
\end{equation}

Using denoising score matching, the true score can be written as:
\begin{equation}
	\nabla_{{\bf x}_t} \log p_t({\bf x}_t | {\bf \hat{y}}) = -\dfrac{{\bf x}_t - \mu({\bf x}_0, {\bf \hat{y}}, t)}{\sigma(t)^2}.
	\label{eq:Score_Mathcing}
\end{equation}

Finally, we define our training loss as
\begin{equation}
	L_{\text{score}} = \arg\min_{\theta} \mathbb{E}_{t, ({\bf x}_0, {\bf \hat{y}}), \boldsymbol{\phi}} \left[ \left\| s_{\theta}({\bf x}_t, {\bf \hat{y}}, \boldsymbol{\bf v}, t) + \frac{\boldsymbol{\phi}}{\sigma(t)} \right\|^2 \right],
	\label{eq:Score_Loss}
\end{equation}
where ${\boldsymbol{\phi}}$ is sampled from a Gaussian distribution with zero mean and identity covariance. 
\subsection{Audio-Visual Fusion}

To effectively leverage complementary information from both audio and visual modalities, we introduce a cross-attention-based fusion mechanism in our framework. Inspired by Stable Diffusion \cite{HighRes}, which demonstrates that cross-attention is effective for incorporating conditional inputs from various modalities in image generation tasks, we extend this idea to the audio-visual domain for SE. Specifically, we integrate visual cues into both the predictive and generative audio models using the cross-attention blocks.

In these cross-attention blocks, the audio features serve as queries, while the keys and values are replaced with visual embeddings obtained from our visual encoder. This design enables the models to selectively attend to relevant visual information, thereby enriching the audio representation. We hypothesize that supplementing the audio input with visual context leads to more accurate speech reconstruction, especially under challenging acoustic conditions. Importantly, this fusion mechanism improves the overall performance of the SE system without incurring additional inference cost.

The audio-visual feature is computed as:
\begin{equation}
	\begin{array}{l}
		{\bf \tilde h} = {\bf CA} \left( {\bf v}, {\bf h} \right),
	\end{array}
	\label{eq:AV_Cross_Attn}
\end{equation}
where ${\bf \tilde h}$ is the resulting audio-visual feature, ${\bf h}$ is the original audio feature from the predictive or generative model, and ${\bf CA}$ denotes the cross-attention operation with visual embedding ${\bf v}$ as the conditioning input.

% ++++++++++++++++++++++++++++++++++++Text Modality++++++++++++++++++++++++++++++++++++
\subsection{Linguistic Data Preparation}
\label{sect:textmodal}
%\hspace{3ex}
In the proposed DLAV-SE system, linguistic information is incorporated through CMKT. When available, the corresponding transcriptions of speech signals can be used directly to obtain this information. However, many datasets do not provide accessible transcriptions. To address this issue and prepare the linguistic input, we may leverage a high-performance ASR system to transcribe each clean speech sample into text. To achieve high recognition accuracy, this study adopts the large version of the Whisper ASR model. \cite{Whisper} in this study.

\subsection{Linguistic Contextual Guidance}
\label{sect:textmodal}
To incorporate high-level semantic context into the DLAV-SE framework, we extract linguistic representations from transcribed speech using a pretrained language model. In the right branch of Fig.~\ref{fig:training_framework}, the context-dependent linguistic features are derived from a pretrained BERT model. These features, referred to as linguistic guidance, are used to inform the audio-visual modality and enhance context-aware SE. The process of extracting linguistic guidance can be formulated as

\begin{equation}
	\begin{array}{l}
		{\bf t}_{{\rm token}}  = \text{Tokenizer}\left( {\bf \tau} \right) \\ 
		\vspace{2mm}
		{\bf Z}_0  = \left[ \text{CLS}, {\bf t}_{{\rm token}}, \text{SEP} \right] \\ 
		\vspace{2mm}
		{\bf Z}  = \text{BERT} \left( {\bf Z}_0 \right), \\ 		
	\end{array}
	\label{eq:bert}
\end{equation}
where ${\bf \tau}$ denotes the text sequence generated by Whisper \cite{Whisper}, and ${\bf t}_{{\rm token}}$ represents the corresponding word-piece-based tokens. The BERT model receives an input sequence ${\bf Z}_0$ that includes special classification (CLS) and separator (SEP) tokens. The resulting output ${\bf Z} \in \mathbb{R}^{T_t \times d_s}$ captures rich, context-dependent linguistic information, where $T_t$ is the length of the tokenized text sequence and $d_s$ is the token embedding dimension.

\subsection{Cross-Modal Knowledge Transfer (CMKT)}
To enable AVSE models to leverage contextual linguistic knowledge, we propose the CMKT mechanism that integrates text-based supervision during training. This mechanism enforces the audio-visual model to encode context-dependent linguistic information by introducing additional losses from the linguistic modality. Specifically, we align the audio-visual features from the left branch with linguistic representations derived from BERT, thereby bridging the modality gap through both alignment and adaptation strategies.

However, a fundamental challenge arises due to the dimensional and distributional differences between the audio-visual features $\mathbf{H} \in \mathbb{R}^{T_a \times d_a}$ and the linguistic features $\mathbf{Z} \in \mathbb{R}^{T_t \times d_t}$. To address this domain mismatch, we adopt a cross-modal alignment strategy along with a cross-modal neural adapter, inspired by Lu et al. \cite{LuCTCOP}.

The cross-modal alignment module consists of a fully connected layer $\mathbf{FC1}$ followed by a feature fusion layer. The purpose of this design is to transfer contextual information from the linguistic domain into the audio-visual latent space. We project the audio-visual feature $\mathbf{H}$ into the latent space of the linguistic embedding using $\mathbf{FC1}$:

\begin{equation}
    \mathbf{\tilde H} = \mathbf{FC1}(\mathbf{H}) \in \mathbb{R}^{T_a \times d_t},
    \label{eq:FC1}
\end{equation}

where $\mathbf{\tilde H}$ is the projected audio-visual latent feature.

We explore two distinct fusion methods for integrating linguistic context: MHCA and OT.

\subsubsection{Multi-Head Cross-Attention (MHCA)}

MHCA fuses audio-visual and linguistic modalities through stacked cross-attention layers. First, the text input is tokenized and embedded with position encoding:

\begin{equation}
    \mathbf{\tilde Z}_L^0 = \text{EMB}(\mathbf{Z}_0) + \mathbf{PE}_T,
    \label{eq:text_emb}
\end{equation}

where $\mathbf{Z}_0$ is the raw text embedding, EMB is the token embedding layer, and $\mathbf{PE}_T$ is positional encoding.

In each cross-attention layer, linguistic features are updated as follows:

\begin{equation}
    \mathbf{\tilde Z}_L^i = \text{MHCA}(\mathbf{\tilde H}, \mathbf{\tilde Z}_L^{i-1}) \in \mathbb{R}^{T_t \times d_t},
    \label{eq:CMKT_Cross_Attn}
\end{equation}

for $i = 1, ..., M_t$. After $M_t$ layers, a layer normalization and linear transformation yield the final cross-modal embedding:

\begin{equation}
    \mathbf{\tilde Z}_f = \text{FC}_H(\text{LN}(\mathbf{\tilde Z}_L^{M_t})) \in \mathbb{R}^{T_t \times d_s}.
    \label{eq:CMKT_Cross_Head}
\end{equation}

\subsubsection{Optimal Transport (OT)}

Alternatively, we adopt an OT-based fusion strategy to align audio-visual and linguistic distributions. OT aims to find a minimal-cost transport plan between the two feature spaces via a projection matrix learned during training:

\begin{equation}
    L_{\text{OPT}}(\mathbf{Z}, \mathbf{H}) = \min_{\gamma \in \Pi(U(\mathbf{Z}, \mathbf{H}))} \sum_{i,j} \gamma(\mathbf{z}_i, \mathbf{h}_j) C(\mathbf{z}_i, \mathbf{h}_j),
    \label{eq:OT_loss}
\end{equation}

where $\gamma$ is the transport plan, and $C(\cdot, \cdot)$ is the cosine distance:

\begin{equation}
    C(\mathbf{z}_i, \mathbf{h}_j) = 1 - \cos(\mathbf{z}_i, \mathbf{h}_j).
    \label{eq:transport_cost}
\end{equation}

The optimal transport matrix $\gamma^*$ is computed using Sinkhorn iterations \cite{SinkAttn, VillanoBook}:

\begin{equation}
    \gamma^0 = \exp(-C/\beta), \quad \gamma^{k+1} = F_c(F_r(\gamma^k)),
    \label{eq:sinkhorn_iteration}
\end{equation}

where $F_r$ and $F_c$ are row and column normalizations:

\begin{equation}
\begin{aligned}
    F_r(\gamma) &= \frac{\gamma}{\sum_j \gamma_{i,j}}, \\
    F_c(\gamma) &= \frac{\gamma}{\sum_i \gamma_{i,j}}.
\end{aligned}
    \label{eq:sinkhorn_normalization}
\end{equation}

After convergence, the final context-aware embedding is obtained as:

\begin{equation}
    \mathbf{\tilde Z}_f = \gamma^* \cdot \mathbf{H} \in \mathbb{R}^{T_t \times d_t}.
\end{equation}

\subsubsection{Cross-Modal Neural Adapter}

To further enhance the generative model's feature space, we reproject the latent feature $\mathbf{\tilde H}$ back to the audio-visual space using a second fully connected layer:

\begin{equation}
    \mathbf{\hat H} = \mathbf{FC2}(\mathbf{\tilde H}) \in \mathbb{R}^{T_a \times d_a}.
    \label{eq:adapter}
\end{equation}

The refined representation $\mathbf{\hat H}$ is added back to the original feature $\mathbf{H}$ to form a context-enriched audio-visual feature:

\begin{equation}
    \mathbf{H}_t = \mathbf{H} + \beta \cdot \mathbf{\hat H},
    \label{eq:adapter_add}
\end{equation}

where $\beta$ controls the influence of linguistic information in the final representation.

\subsubsection{Training Objectives}

To enforce the encoding of linguistic information into the audio-visual modality, we define a cross-modal alignment loss based on cosine similarity:

\begin{equation}
    L_{\text{align}} = 1 - \cos(\mathbf{Z}, \mathbf{\tilde Z}_f).
    \label{eq:align}
\end{equation}

The total training loss differs based on the chosen fusion mechanism:

\paragraph{MHCA}
\begin{equation}
    L_{\text{MHCA}} = \omega \cdot L_{\text{denoiser}} + (1 - \omega) \cdot L_{\text{score}} + \alpha \cdot L_{\text{align}},
    \label{eq:total_loss_MHCA}
\end{equation}

\paragraph{OT}
\begin{equation}
    L_{\text{OT}} = \omega \cdot L_{\text{denoiser}} + (1 - \omega) \cdot L_{\text{score}} + \alpha \cdot (L_{\text{align}} + L_{\text{OPT}}).
    \label{eq:total_loss_OT}
\end{equation}

During inference, only the left branch (generative model) is retained, ensuring both computational efficiency and linguistically informed enhancement.

\section{Experiments}
\label{sec:exp}
%\hspace{3ex}
This section describes the implementation of the proposed AVSE system, incorporating the CMKT process, and presents a comprehensive performance evaluation. We compare our model with several benchmark systems using two standardized datasets to assess its effectiveness. In addition, objective evaluations and subjective listening tests are conducted on a real-world dataset to evaluate the perceptual quality of the enhanced speech.
\subsection{Dataset}
%\hspace{3ex}
Our experiments are conducted on three datasets: the TMSV dataset, the 3rd AVSE Challenge dataset \cite{AVSEChallenge}—referred to as AVSEC-3 in the following discussion—and a realistic in-vehicle dataset recorded in the Academia Sinica Campus, denoted as "AS-CarSpeech" The TMSV dataset comprises speech recordings from 18 native Mandarin speakers (13 males and 5 females), each articulating 320 Mandarin sentences composed of 10 Chinese characters. The duration of each utterance ranges from approximately 2 to 4 seconds. To ensure consistency and reproducibility, we followed the noise injection and train-test split protocols described in \cite{ImprovedLite}. The AVSEC-3 dataset includes speech from 37,823 speakers, with each video clip ranging from 2 to 10 seconds in duration. The AS-CarSpeech dataset consists of 100 realistic recordings captured in car-driving scenarios, each approximately 5 seconds long, featuring speakers uttering single sentences composed of 10 Chinese characters.

\subsection{Model Implementation}
%\hspace{3ex}
To ensure effective multimodal SE, we carefully design preprocessing pipelines and model configurations for each modality, as well as for training and inference procedures. For visual feature extraction, we employ the Mediapipe framework \cite{Mediapipe} to detect facial landmarks and extract an ROI centered on the mouth. Each frame is cropped to a resolution of 96 × 96 pixels and normalized using the mean and standard deviation calculated from the training set. For audio preprocessing, we follow the settings in \cite{SGMSE}. Audio signals are resampled to 16 kHz and converted into complex-valued STFT representations using a window size of 510 and a periodic Hann window, resulting in a frequency dimension of $F = 256$. The hop length is set to 128 for TMSV and 160 for AVSEC-3. Each STFT spectrogram is truncated to $T = 256$ time frames, enabling random cropping during batch training.

For the predictive model, we adopt a modified version of NCSN++ from \cite{ScoreDiff}. The model is used in a deterministic setting by fixing the timestep to 1, and its output is treated as a preliminary denoised speech spectrogram. The underlying UNet architecture follows the original implementation in \cite{SGMSE}. For the generative model, we also adapt NCSN++ from \cite{SGMSE} as the backbone. The stochastic diffusion process is parameterized with $\boldsymbol{\sigma}_{\text{min}} = 0.05$, $\boldsymbol{\sigma}_{\text{max}} = 0.5$, and $\eta = 1.5$. During training, we track an exponential moving average of the model weights with a decay factor of 0.999, which is used during inference. All input spectrograms $({\bf x_t}, {\bf \hat{y}})$ are represented in the complex STFT domain. To accommodate the real-valued NCSN++ model, the real and imaginary parts are separated into distinct channels before input. These channels are later recombined into complex-valued outputs for loss computation. In the inference stage, we employ a Predictor-Corrector (PC) sampler \cite{ScoreDiff} with 30 reverse diffusion steps. The corrector utilizes annealed Langevin dynamics (ALD) \cite{ALD} with a single correction step per iteration.

After downsampling in the generative model, the audio-visual feature dimension becomes $(B, C, F, T) = (B, 256, 4, 4)$. To align dimensions between modalities, we apply a fully connected layer $\mathbf{FC1}$ with shape $768 \times 4$, and a reverse transformation $\mathbf{FC2}$ with shape $4 \times 768$. For the MHCA variant, we use $M_t = 6$ MHCA layers, each with 4 attention heads. The loss coefficients in Equation (\ref{eq:total_loss_MHCA}) are set to $\omega = 0.5$ and $\alpha = 0.2$. For the OT variant, we use $\omega = 0.5$, $\alpha = 0.01$ in Equation (\ref{eq:total_loss_OT}), and $\beta = 0.5$ in Equation (\ref{eq:sinkhorn_iteration}). The weighting coefficient for the cross-modal neural adapter is set to $\beta = 0.1$ in Equation (\ref{eq:adapter_add}). For training, we use the Adam optimizer with a learning rate of $10^{-4}$ and a batch size of 1.

%\vspace{-4mm}
\subsection{Objective Evaluation Results}
%\hspace{3ex}
\label{sec:results}
This subsection evaluates the effectiveness of incorporating linguistic and visual modalities in our proposed DLAV-SE system and demonstrates the advantages of our CMKT design during training. We also evaluate the DLAV-SE system using a single modality (audio-only) and compare it with its multimodal variants: audio-visual and audio-visual-linguistic.

We first assess the contribution of the visual and linguistic modalities to the SE task. These modalities are expected to provide auxiliary cues that support the audio model in learning more robust representations. We compare three model variants on the TMSV dataset: (1) DLAV-SE with audio-only input (“A”), (2) DLAV-SE with audio and linguistic inputs (“A+L”), and (3) DLAV-SE with all three modalities (“A+V+L”). The results are presented in Table \ref{tab0}.
\begin{table}
    \caption{Objective testing results of SE with different modalities on the TMSV dataset}
    \centering
    \begin{tabular}{lcccc}
        \toprule
        & \multicolumn{3}{c}{} \\ %\cmidrule(lr){5-5}
        Method & Modality & PESQ${\tiny \uparrow}$  & STOI${\tiny \uparrow}$ & SI-SDR${\tiny \uparrow}$ \\ 
        \midrule
        Noisy & - & 1.19 & 0.60 & -5.5 \\
       DLAV-SE & A  & 1.54 & 0.69 & 1.4 \\
        DLAV-SE & A+L  & 1.66 & 0.70 & 1.9 \\
        \textbf{DLAV-SE} & \textbf{A+V+L}  & \textbf{1.74} & \textbf{0.72} & \textbf{4.1} \\ 
        \bottomrule
    \end{tabular}
    \vspace{2mm}
    \label{tab0}
\end{table}
In Table \ref{tab0}, “Noisy” denotes the original noisy input. “A” refers to our audio-only baseline, trained without visual embeddings or CMKT. “A+L” incorporates linguistic loss via CMKT but omits visual information. “A+V+L” represents the proposed DLAV-SE framework. These results show that adding linguistic information improves performance over the audio-only model, and including visual cues further enhances SE quality, achieving the best performance across all metrics.

We next compare the DLAV-SE system to several SOTA AVSE benchmarks on the TMSV and AVSEC-3 datasets. Results are shown in Tables \ref{tab1} and \ref{tab2}, respectively. Performance is evaluated using perceptual evaluation of speech quality (PESQ) \cite{PESQ}, short-time objective intelligibility (STOI) \cite{STOI}, and scale-invariant signal-to-distortion ratio (SI-SDR) \cite{SISDR}. PESQ and STOI assess speech quality and intelligibility, respectively, while SI-SDR measures signal-level distortion; all are intrusive objective metrics requiring clean reference, with higher scores indicating better performance. 

We also indicate the input modalities used in each system.

\begin{table}
    \caption{Performance comparison of representative benchmark AVSE systems on the TMSV dataset}
    \centering
    \begin{tabular}{lcccc}
        \toprule
        & \multicolumn{3}{c}{} \\ %\cmidrule(lr){5-5}
        Method & Modality & PESQ${\tiny \uparrow}$  & STOI${\tiny \uparrow}$ & SI-SDR${\tiny \uparrow}$ \\ 
        \midrule
        Noisy & - & 1.19 & 0.60 & -5.5 \\
        AVCVAE \cite{AVSE6} & A+V & 1.34 & 0.63  & - \\
        SSL-AVSE \cite{SSLAVSE} & A+V & 1.41 & 0.68 & - \\
        ILAVSE \cite{ImprovedLite} & A+V & 1.41 & 0.64 & - \\
        DCUC-Net \cite{DCUC-Net} & A+V & 1.41 & 0.66 & - \\
        \midrule
        DLAV-SE   & A+V & 1.74 & 0.71  & 2.5  \\
        \textbf{DLAV-SE (+MHCA)} & \textbf{A+V+L} & \textbf{1.74} & \textbf{0.72} & \textbf{4.1} \\
        DLAV-SE (+OT) & A+V+L & 1.66 & 0.69 & 2.5 \\ 
        \bottomrule
    \end{tabular}
    \vspace{2mm}
    \label{tab1}
\end{table}

\begin{table}
    \caption{Performance comparison of representative benchmark AVSE systems on the AVSEC-3 dataset}
    \centering
    \begin{tabular}{lcccc}
        \toprule
        & \multicolumn{3}{c}{} \\%\\ \cmidrule(lr){3-5}
        Method & Modality & PESQ${\tiny \uparrow}$ & STOI${\tiny \uparrow}$ \\ 
        \midrule
        Noisy & - & 1.46 & 0.61 \\ 
        Baseline & A+V & 1.49 & 0.62 \\
        LSTMSE-Net \cite{LSTMSE-Net} & A+V & 1.55 & 0.65 \\
        MMDTN \cite{MMDTN} & A+V & 1.73 & 0.69 \\
        \midrule
        DLAV-SE  & A+V & 1.91 & 0.68 \\ 
       DLAV-SE (+MHCA) & A+V+L & 1.82 & 0.66 \\
        \textbf{DLAV-SE (+OT)} & \textbf{A+V+L} & \textbf{1.94} & \textbf{0.69} \\ 
        \bottomrule
    \end{tabular}
    \vspace{2mm}
    \label{tab2}
\end{table}
Here, “DLAV-SE (+MHCA)” refers to the DLAV-SE model with MHCA, while “DLAV-SE (+OT)” includes the OT-based alignment. On the TMSV dataset (Table \ref{tab1}), DLAV-SE (+MHCA) achieves SOTA results across all metrics. On AVSEC-3 (Table \ref{tab2}), DLAV-SE (+OT) outperforms the AVSE Challenge baseline and existing benchmarks. These findings validate the effectiveness of our CMKT framework in facilitating linguistic knowledge transfer, ultimately boosting audio-visual SE performance. 

To further evaluate the robustness of the proposed DLAV-SE system, we conducted experiments using the AS-CarSpeech dataset. As this dataset does not include clean reference signals, the previously mentioned intrusive metrics could not be applied. Instead, we employed four non-intrusive objective metrics that do not require clean references to assess the system's performance. The selected metrics are DNSMOS \cite{DNSMOS}, MOSA-Net+ \cite{MOSA-Net, MOSA-Net+}, SpeechLMScore \cite{SpeechLMScore}, and VQScore \cite{VQScore}. DNSMOS provides three sub-scores: Overall Quality (OVRL), Signal Quality (SIG), and Background Noise Quality (BAK). MOSA-Net+ estimates both speech quality and intelligibility from a perceptual perspective. SpeechLMScore measures the linguistic plausibility of enhanced speech based on perplexity derived from a pretrained speech language model. VQScore \cite{VQScore} evaluates perceptual quality using vector-quantized self-supervised representations.  The evaluation results of the DLAV-SE system using different modality combinations—namely “A”, “A+V”, and “A+V+L”, are summarized in Table \ref{tab:merged_quality}. The experimental results in Table \ref {tab:merged_quality} clearly indicate that the full multimodal configuration (A+V+L) significantly enhances speech quality across all evaluation metrics. This setup significantly outperforms the other configurations, attaining the highest OVRL (3.18), SIG (3.55), MOSA-Net+ Quality (4.50), Intelligibility (0.997), and VQScore (0.704). It also achieves the lowest SpeechLMScore (1.613), indicating the greatest linguistic plausibility. These results underscore the importance of leveraging linguistic context to generate more natural and coherent speech outputs.

Overall, the findings affirm the efficacy of the proposed DLAV-SE framework in challenging real-world conditions. The integration of multimodal inputs, particularly the inclusion of linguistic features, substantially enhances both the perceived quality and intelligibility of the enhanced speech. This demonstrates the potential of audio-visual-linguistic fusion for robust speech enhancement in noisy environments.

\begin{table*}[t]
    \caption{Non-intrusive objective testing results of SE with different modalities on the AS-CarSpeech dataset}
    \centering
    \setlength{\tabcolsep}{12pt} % Adjust column spacing (default is 6pt)
    \begin{tabular}{ll|ccc|cc|c|c}
        \toprule
        \multicolumn{2}{c|}{\textbf{Method}} & \multicolumn{3}{c|}{\textbf{DNSMOS}} & \multicolumn{2}{c|}{\textbf{MOSA-Net+}} & \textbf{SpeechLMScore} & \textbf{VQScore} \\
        \cmidrule(lr){1-2} \cmidrule(lr){3-5} \cmidrule(lr){6-7} \cmidrule(lr){8-8} \cmidrule(lr){9-9}
        \textbf{Name} & \textbf{Modality} & OVRL${\tiny \uparrow}$ & SIG${\tiny \uparrow}$ & BAK${\tiny \uparrow}$ & Quality${\tiny \uparrow}$ & Intelligibility${\tiny \uparrow}$ & Score${\tiny \downarrow}$ & Score${\tiny \uparrow}$ \\
        \midrule
        Noisy & - & 2.76 & 3.52 & 3.12 & 4.36 & 0.995 & 1.794 & 0.665 \\
        DLAV-SE & A & 2.49 & 2.76 & \textbf{3.94} & 3.62 & 0.954 & 1.729 & 0.656 \\
        DLAV-SE & A+V & 2.82 & 3.15 & 3.89 & 4.22 & 0.991 & 1.730 & 0.674 \\
        \textbf{DLAV-SE} & \textbf{A+V+L} & \textbf{3.18} & \textbf{3.55} & 3.92 & \textbf{4.50} & \textbf{0.997} & \textbf{1.613} & \textbf{0.704} \\
        \bottomrule
    \end{tabular}
    \label{tab:merged_quality}
\end{table*}

\subsection{Spectrogram Analysis}
\begin{figure}[tb]
	\centering
    \includegraphics[scale=0.17]{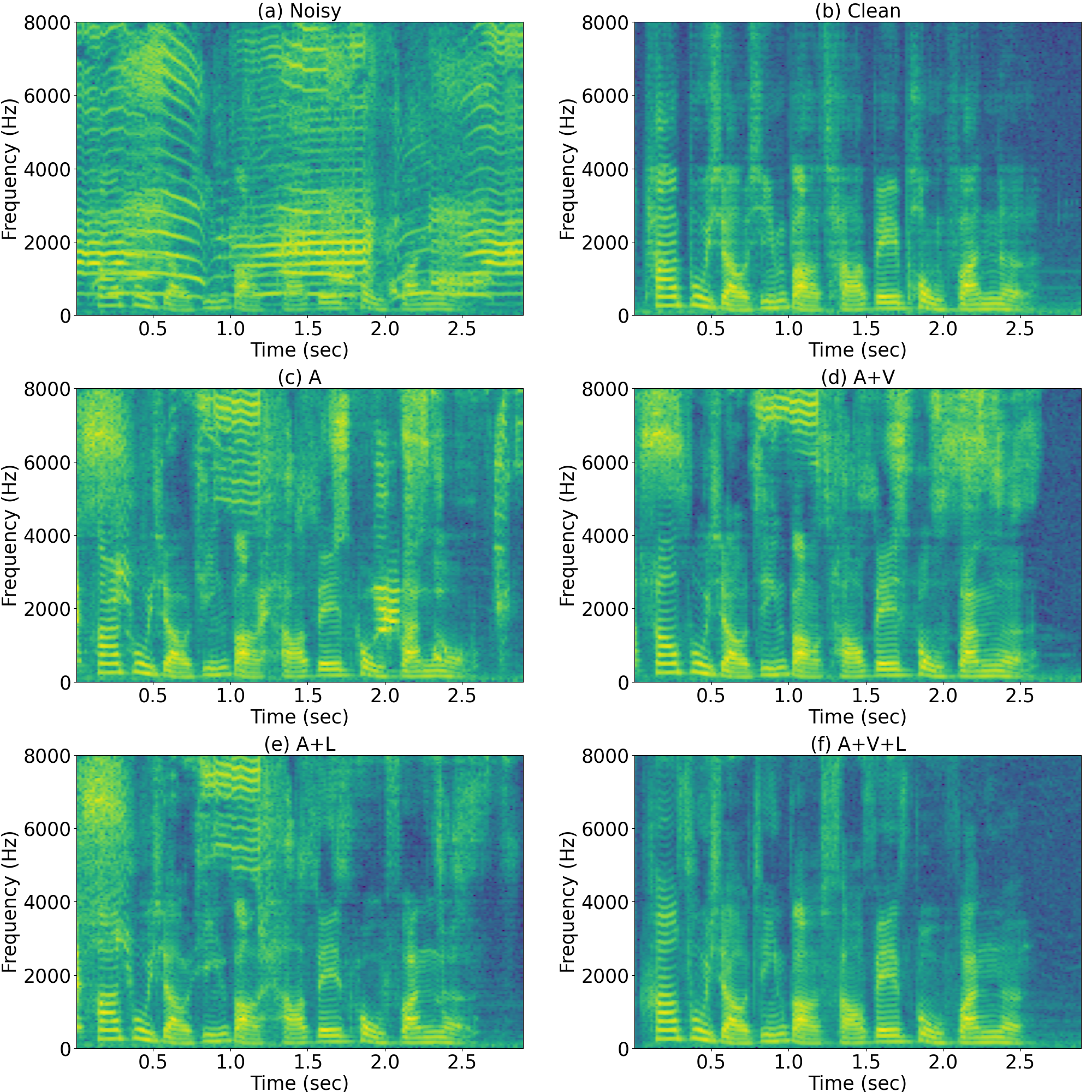} 
    \caption{Spectrograms of (a) noisy speech, (b) clean speech, (c) enhanced speech by our model ``DLAV-SE (A)", (d) enhanced speech by our model ``DLAV-SE (A+V)", (e) enhanced speech by our model ``DLAV-SE (A+L)", (f) enhanced speech by our model ``DLAV-SE (A+V+L)" for an example in the TMSV dataset. The vertical axis represents frequency, and the horizontal axis represents time.}
	\vspace{-4mm}
    \label{fig:TMSV_spec}
\end{figure}

To further explore the impact of the linguistic modality, we conduct spectrogram analysis on an example from the TMSV dataset. Fig. \ref{fig:TMSV_spec} illustrates spectrograms of the noisy input, clean target, and enhanced outputs from models using various modality combinations.
At the end of the clean spectrogram, low-frequency energy drops sharply as the speaker stops talking. The audio-only model (“A”) fails to replicate this pattern accurately. In contrast, “A+L” shows better suppression of residual noise, suggesting that linguistic tokens help distinguish between pure noise and noisy speech through contextual cues. It is also noted that the clean spectrogram exhibits reduced high-frequency energy, consistent with the nature of human speech. Comparing “A+V” and “A+V+L”, we find that the inclusion of linguistic input further improves noise suppression, especially in high-frequency regions. This implies that textual features support the reconstruction of cleaner speech signals by offering additional semantic guidance.

\subsection{Subjective Listening Test}
To assess the effectiveness of our proposed method from a human perceptual standpoint, we conducted a subjective listening test using the AS-CarSpeech dataset. A total of 50 participants (23 male, 27 female), aged between 20 and 50, were recruited for the evaluation. \footnote{All ethical and experimental procedures were approved by the Institutional Review Board (IRB) on Biomedical Science Research at Academia Sinica under protocol number AS-IRB-BM-24026. All experiments were conducted in accordance with the approved IRB protocols.} Each participant was presented with 100 speech samples randomly selected from a pool of 300 utterances, which included 100 recordings processed by each of the three compared methods. Participants rated each sample based on two criteria: Quality: defined as speech clarity and background noise suppression, scored on a 5-point Likert scale (1–5); intelligibility: measured by the number of correctly recognized characters from 10 words. We focus on investigating the two best-performing SE systems from the previous experiments, namely 'A+V' and 'A+V+L'. The results are summarized in Table \ref{tab5}.

\begin{table}
    \caption{Subjective Quality and Intelligibility of SE with different modalities on the AS-CarSpeech dataset}
    \centering
    \begin{tabular}{lcccc}
        \toprule
        & \multicolumn{3}{c}{} \\%\\ \cmidrule(lr){3-5}
        Method & Modality & Quality${\tiny \uparrow}$ & Intelligibility${\tiny \uparrow}$ \\ 
        \midrule
        Noisy  & - & 3.60 & 0.949 \\ 
        DLAV-SE  & A+V & 3.80 & 0.953\\
        \textbf{DLAV-SE} & \textbf{A+V+L} & \textbf{4.31} & \textbf{0.964} \\ 
        \bottomrule
    \end{tabular}
    \vspace{2mm}
    \label{tab5}
\end{table}

In Table \ref{tab5}, “Noisy” refers to the original unprocessed audio input, “A+V” denotes the DLAV-SE model trained without linguistic guidance, and “A+V+L” corresponds to our full DLAV-SE framework incorporating CMKT. The results indicate that our AVSE system consistently improves both perceived speech quality and intelligibility. Notably, the inclusion of linguistic modality through CMKT leads to a significant further enhancement, underscoring the benefit of linguistic knowledge transfer from a human perceptual perspective. These subjective findings align well with the objective evaluation results discussed in earlier subsections.

\subsection{Discussion}
%\hspace{3ex}
Several directions remain open for further investigation. Our experiments indicate that the hyperparameters within the CMKT framework play a critical role in performance, particularly the weights assigned in the loss function. To better understand their impact, we plan to conduct ablation studies focusing on two key parameters: the weight of the cross-modal alignment loss ($\alpha$) and the weight of the cross-modal neural adapter ($\beta$). These hyperparameters govern the degree of domain gap reduction and are instrumental in facilitating effective linguistic knowledge transfer. In addition, we aim to evaluate the generalizability of our approach by testing it on broader benchmark datasets such as VoxCeleb2 and LRW. Finally, recognizing that the incorporation of linguistic guidance increases model complexity and computational cost, we intend to explore more lightweight and efficient implementations of the linguistic modality in future work.

\section{Conclusion}
\label{sec:conclusion}
%\hspace{3ex}
In this study, we present DLAV-SE, a novel diffusion-model-based AVSE system that integrates linguistic information into the audio-visual framework. Our approach builds upon a score-based diffusion model comprising predictive and generative stages, achieving competitive performance on three AVSE datasets. To incorporate linguistic modality, we introduce two CMKT mechanisms: MHCA and OT. These components bridge the domain gap between modalities and leverage the pretrained BERT language model to transfer contextual linguistic knowledge into the audio-visual domain. Importantly, this integration enhances performance during training without introducing additional inference overhead.

Our experiments demonstrate the effectiveness of the proposed system on datasets spanning different languages, suggesting that the method is broadly generalizable across linguistic contexts. Furthermore, a subjective listening test conducted on a realistic dataset validates the perceptual benefits of CMKT from a human listener's perspective. In future work, we aim to develop a more compact and efficient implementation of the AVSE system and explore broader applications within multi-modal SE frameworks.

% or


%\appendices

%\section{}

%\section*{Acknowledgment}


\ifCLASSOPTIONcaptionsoff
  \newpage
\fi

\end{document}